%% file: main.tex
\documentclass[preprint,12pt]{elsarticle}

\usepackage{amssymb}

\usepackage{comment}
\usepackage{tabularx}
\journal{}

\begin{document}

\begin{frontmatter}



\title{The Influence Operation Ontology (IOO)\\\footnotesize{\textit{-Working version 1.0-}}}

\author[inst1]{Alejandro David Cayuela Tudela}

\author[inst2]{Javier Pastor-Galindo}

\author[inst1]{Pantaleone Nespoli}

\author[inst1]{José Antonio Ruipérez-Valiente}

\affiliation[inst1]{organization={Department of Information and Communications Engineering, University of Murcia},
            postcode={30100}, 
            country={Spain}}

\affiliation[inst2]{organization={Computer Systems Engineering Department, Universidad Politecnica de Madrid}, 
             postcode={28031}, 
             country={Spain}}

\begin{abstract}
Ontologies provide a systematic framework for organizing and leveraging knowledge, enabling smarter and more effective decision-making. In order to advance in the capitalization and augmentation of intelligence related to nowadays cyberoperations, the proposed Influence Operation Ontology establishes the main entities and relationships to model offensive tactics and techniques by threat actors against the public audience through the information environment. It aims to stimulate research and development in the field, leading to innovative applications against influence operations, particularly in the fields of intelligence, security, and defense.
\end{abstract}




\end{frontmatter}


\section{Introduccion}
\label{sec:intro}

According to the Global Risk Report 2025 by the World Econonomic Forum, Misinformation and disinformation represent the top one of the threats in the short term~\cite{worldForum2025}. In addition, the polarization of society is risk number four. Malicious actors are using these threats to menace the integrity of nations by manipulating public perception and influencing citizens~\cite{Pamment2020}. 

The coordination of all these efforts to undermine the integrity of societies by employing deceptive and illegitimate tactics with altering and manipulating the population is called Influence Operations (IOs)~\cite{Pamment2020}. According to European cybersecurity institutions, IOs rank among the ten most prevalent and significant threats in the region~\cite{enisa2024,1eeas_report,2eeas_report}. In response, Europe is actively working to establish a common framework for analyzing these threats~\cite{Pamment2020,disarm_framework} and equipping states with effective countermeasures~\cite{2eeas_report}.

However, the multidisciplinary nature of IOs makes it particularly challenging to characterize the information environment and its key components, such as the channels where attacks unfold, the communities that emerge within online networks, and the narratives that gain traction~\cite{pastor2025influence}. 
This work introduces an ontology for influence operations that allows capturing the multiple domains that compose the information environment. It is an approach focused on cyber threat intelligence (CTI), facilitating interoperability with languages and CTI sharing platforms~\cite{gonzalez2025cti}. In addition, it is developed using as a base the most understood frameworks as well as the most important proposals, allowing to unify knowledge in a single analysis tool, improving the analysis and information sharing capabilities in a standardized way.

\input{ontology_objects}



\bibliographystyle{elsarticle-num} 
\bibliography{cas-refs}





\newpage
\appendix
\input{appendix}
\label{appendix}

\end{document}

%% file: ontology_objects.tex
\section{Influence Operation Ontology specification }\label{sec:ontology_objects}

Although various perspectives could be adopted, our ontology focuses on the intersection of cyber threat intelligence (CTI) and influence operations with an additional socio-technical aspect. This approach allows for representing the traditional components of cyber intelligence, such as threat actors or attack vectors, by integrating them with the social context and information ecosystem in which the operations occur.  This is why the ontology is inspired by the well-known STIX v2.1~\cite{stix} language widely used in CTI sharing, the DISARM framework~\cite{disarm_framework}, ABCDE framework~\cite{Pamment2020} and the extensions proposed by Filigran~\cite{filigran}. STIX provides a standardized and extensible language that CTI platforms understand, DISARM the specification of malicious actor sharing, ABCDE is a framework that breaks down the disinformation problem into small operational factors that can be addressed as questions, and finally, the STIX v2.1 extension proposed by Filigran that adopts new terms for modeling information environments. 

\begin{figure*}[h]
\centerline{
\includegraphics[width=1.5\textwidth]{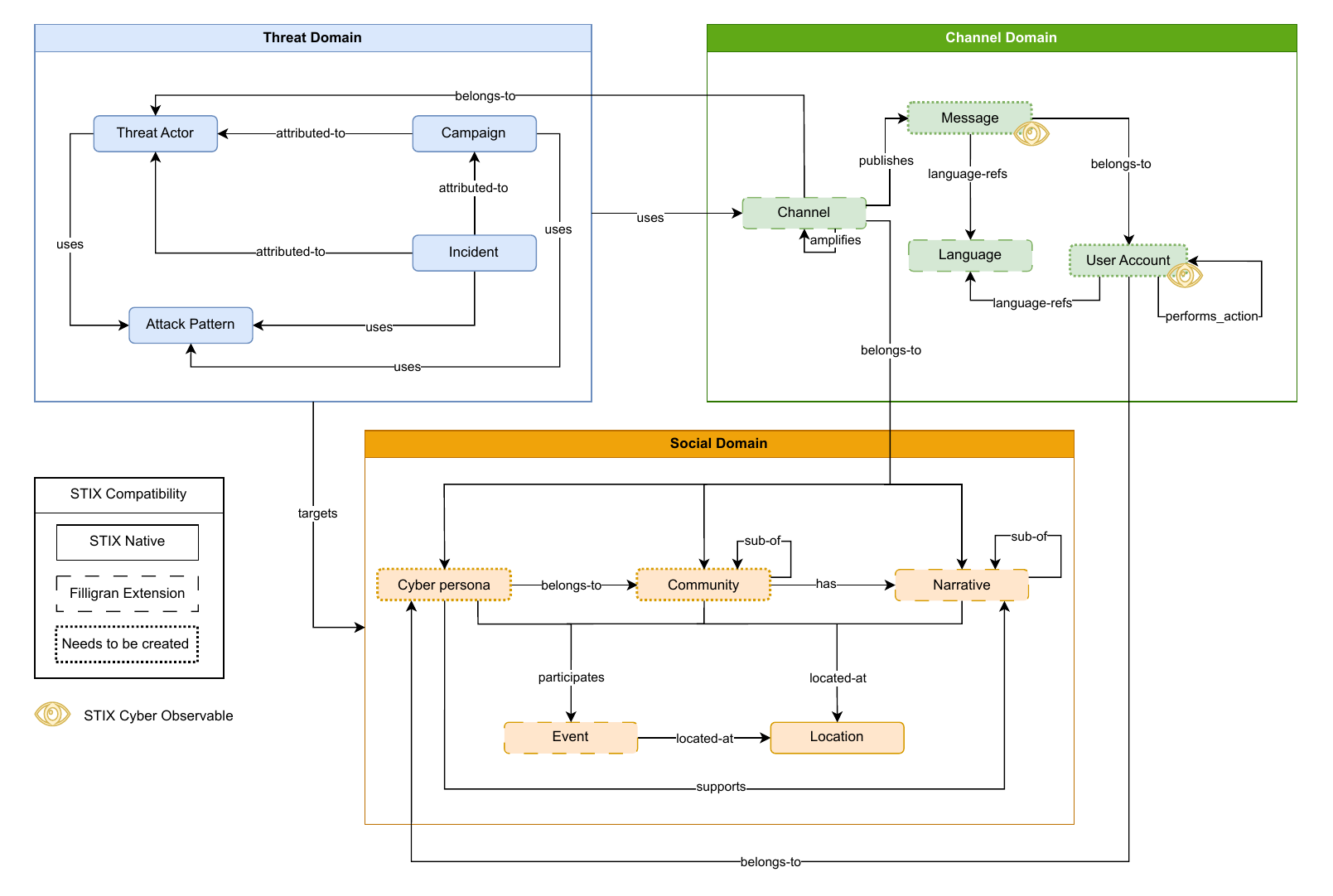}
}
\caption{Influence operations ontology}
\label{fig:ontology}
\end{figure*}

An IO encompasses a variety of actors (malicious or targeted), the information ecosystem, and social structures. The proposed ontology comprises three domains (\figurename~\ref{fig:ontology}) that simplify and facilitate the understanding and characterizing the components of IOs. The Threat domain (Section~\ref{sub:threat-domain}) represents threat actors coordinating efforts and attacks against a specific target. These actors (threats and targets) share a common medium called information ecosystem or Channel domain (Section~\ref{sub:channel-domain}), which both interact. Finally, the Social domain (Section~\ref{sub:social-domain}) defines the actors that use (or live in) the information environment daily and that are susceptible to attack. It also characterizes who is, what is, or where is the target of IOs. These actors (and possible objectives) include people, communities, narratives, events, and locations.

\subsection{Threat domain}\label{sub:threat-domain}
\begin{figure*}[h]
\centerline{
\includegraphics[width=1\textwidth]{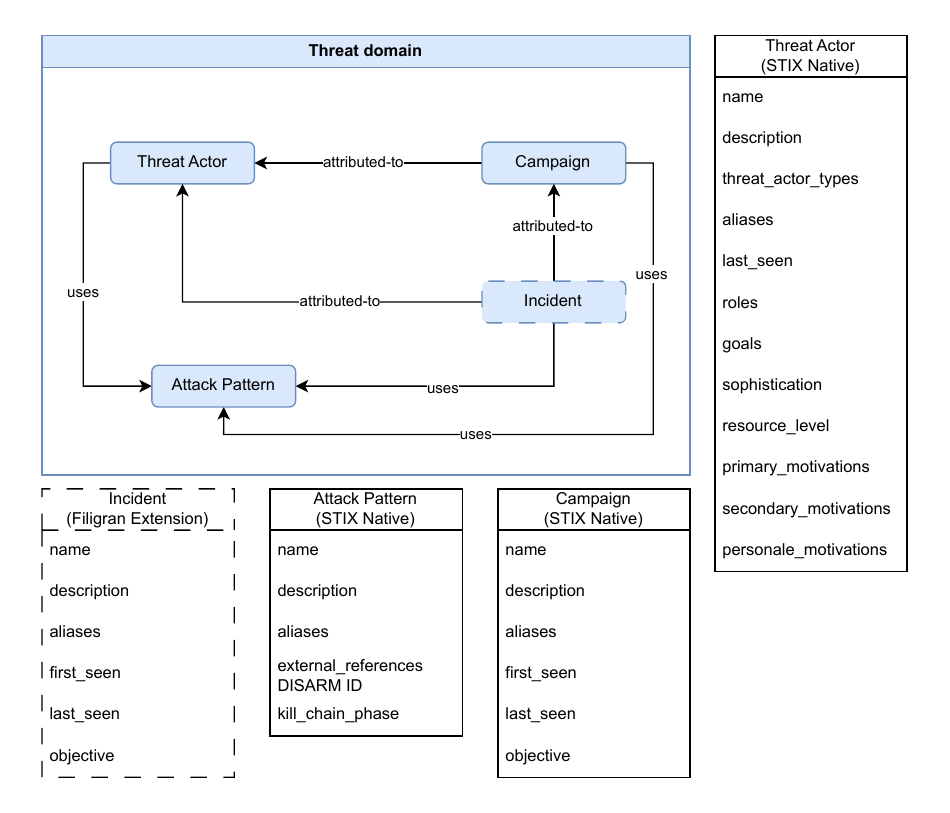}
}
\caption{Threat domain visual representation}
\label{fig:threat-domain}
\end{figure*}

In the context of IOs, threats represent the actors, the collection of incidents, and the particular actions whose objective is to manipulate a \textit{target} perceptions, attitudes, and behaviors~\cite{Pamment2020,1eeas_report,fbiCisa2024}. These threats may come from nation-states, organized groups, or individuals to influence public opinion, destabilize societies, or affect political and economic processes \textit{using} information channels~\cite{1eeas_report,fbiCisa2024}. It is noteworthy that the classes \texttt{Incident}, \texttt{Attack Pattern},  \texttt{Campaign}, and \texttt{Threat Actor} are imported from STIX Version 2.1~\cite{stix} with some adaptations to information environments. The \figurename~\ref{fig:threat-domain} shows all the classes and their relationships in the Threat Domain. 

\subsubsection{\texttt{Incident}}
\label{sec:incident}

An \texttt{Incident} is the group of actions carried out by the \texttt{Threat Actors} trying to accomplish an objective and/or produce a desired effect on the \texttt{Targets}. It is composed of a combination of \texttt{Attack Patterns} that it \textit{uses} and observables~\cite{1eeas_report}. An \texttt{Incident} is defined by its name, description, timestamp, and objective. The complete list of attributes of the \texttt{Incident} class are in \ref{appendix} (Table \ref{tab:incident}).

\subsubsection{\texttt{Attack Pattern}}
\label{subsec:attack pattern}

The \texttt{Attack Pattern} describes how \texttt{Threat Actors} try to influence or manipulate a target audience. \texttt{Attack Patterns} are usually defined by Tactics, Techniques, and Procedures (TTPs). Tactics are the high-level description of the behavior, strategy, and goals of the attack. Techniques are the actions through which threat actors try to accomplish the objective of a tactic. Procedures are the specific (low-level) combination of tasks, techniques, and tactics to conduct an attack and may be unique for different threat actors~\cite{nistcti,1eeas_report}. The attributes defined for \texttt{Attack Pattern}, such as name, description, and kill Chain Phase, are described in \ref{appendix} (Table \ref{tab:attack_pattern}). In the context of IOs, the DISARM TTPS~\cite{disarm_framework} could be encapsulated in this object.

\subsubsection{\texttt{Campaign}}

A \texttt{Campaign} is defined by a name, a description, and an objective. The \texttt{Incidents} carried out over a period of time against specific \texttt{Targets} in a coordinated way could be grouped to the same \texttt{Campaign}. Usually, \texttt{Campaigns} are \textit{attributed to} \texttt{Threat Actors}, stating that they are carrying out the campaign. The attributes defined for \texttt{Campaign} are described in \ref{appendix} (Table \ref{tab:campaign}).

\subsubsection{\texttt{Threat Actor}}

\texttt{Threat actors} are individuals, groups, or organizations believed to be operating intending to modify a \textit{target} audience's perceptions, attitudes, and behaviors. \texttt{Threat actors} are characterized by a name, a description, a threat actor type, goals,  motivations, and more.  They coordinate (\textit{attributed-to}) influence \texttt{Campaigns} and \texttt{Incidents} \textit{using} \texttt{Attack Patterns} to achieve its goals. The complete list of attributes of \texttt{Threat actors} is described in \ref{appendix} (Table \ref{tab:threat_actor}).


\subsection{Channel Domain}\label{sub:channel-domain}

\begin{figure*}[h]
\centerline{
\includegraphics[width=1\textwidth]{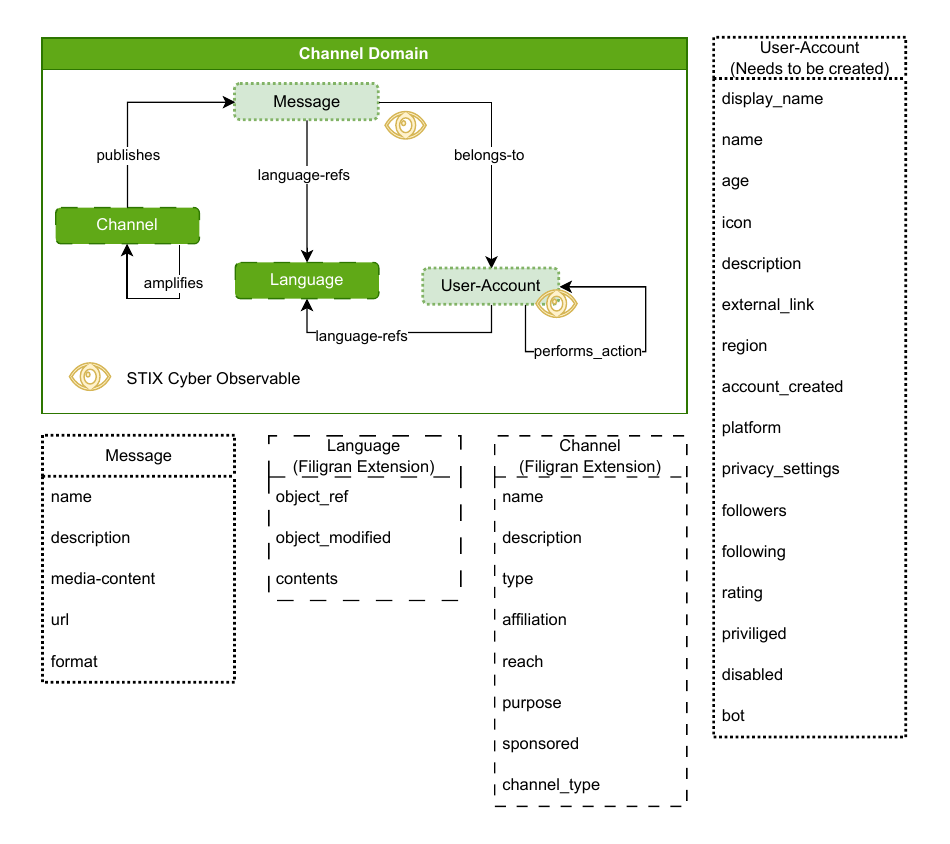}
}
\caption{Channel domain visual representation}
\label{fig:channel-domain}
\end{figure*}

Lasswell proposed the communication model, analyzing some questions regarding how communication works~\cite{lasswell1948}. The base communication model evolved to answer the five following questions: Who/says what/in what channel/to whom/with what effect?~\cite{lasswell1979propaganda}. Currently, this model is one of the most influential and extended in the realm of the media landscape~\cite{lasswelreview}. This work uses this model to characterize the ``Channel Domain''. \figurename~\ref{fig:channel-domain} represents the visual characterization of the actors and their relationships involved in the Channel Domain concerning IOs. The \texttt{User Account} is a class imported from STIX version 2.1 \cite{stix}, while \texttt{Language} and \texttt{Channel} were proposed by Filigran as an extension, and finally, \texttt{Channel} was presented in this ontology. All these classes have been defined or extended in this work.
Additionally, it is necessary to highlight that this domain contains two classes (\texttt{User Account} and \texttt{Message}) that are cyber observables, which are real data that it is possible to see and extract from internet platforms. Cyber observables are elements that help to explain how an \texttt{Incident} occurs~\cite{1eeas_report}. 

\subsubsection{\texttt{User Account}}

A \texttt{User Account} is modeled by a display name, icon, region, and specific attributes according to the platform where the \texttt{User Account} operates as followers, following, rating, or privileged. The \texttt{User Account} represents the presence of a \texttt{Cyber Persona} or a \texttt{Community} within a specific internet platform. The \texttt{User Account} is the owner of the \texttt{Messages} published and could be used by the \texttt{Threat Actors} as channels to spread its influence and manipulate. The complete attributes defined for \texttt{User Account} are described in \ref{appendix} (Table \ref{tab:user_account}).

\subsubsection{\texttt{Channel}}

The \texttt{Channel} is the medium used to \textit{publish} messages from a sender to a receptor. A \texttt{Channel} is characterized by a name, a description, type, affiliation, and purpose. \texttt{Channels} (like websites, social media profiles, groups, or pages) are \textit{used} to send, spread, and \textit{amplify} content or another channel by the \texttt{Threats} to boost the impact of the IOs~\cite{2eeas_report}. The complete list of attributes defined for \texttt{Channel} are described in \ref{appendix} (Table \ref{tab:channel}).

\subsubsection{\texttt{Message}}

The \texttt{Message} corresponds with the information sent from the sender to the receptor with the intent to produce some effect. A \texttt{Message} is defined by a name, a description, the media content, and a format. The complete description of the \texttt{Message} is present in \ref{appendix} (Table \ref{tab:message}). As previously mentioned, it represents a cyber observable being that in IOs a \texttt{Message} can be the data of a post, an article on a website, a YouTube video, an advertisement, etc.


\subsection{Social domain}\label{sub:social-domain}

The Social Domain in the IOs context encompasses the real-world entities projected within the information ecosystem. All these actors are susceptible to becoming targets of IOs, extending the impact not only to individuals but also to entire communities, belief systems, public opinion, and more~\cite{MarshalGanz,LikeWar}. The focal points of influence operations (targets) are those that threat actors seek to influence, manipulate, or alter~\cite{1eeas_report,2eeas_report}. They are modeled through the who, what, and where of the objective and can include \texttt{Cyber Personas}, \texttt{Communities}, or \texttt{Narratives}, as well as specific \texttt{Locations} or \texttt{Events}~\cite{1eeas_report,2eeas_report}. It is necessary to noteworthy that the \texttt{Location} class is imported from STIX version 2.1 \cite{stix}, while \texttt{Narrative} and \texttt{Event} classes were proposed by Filigran \cite{filigran} as an extension, and finally, \texttt{Cyber Persona} and \texttt{Community} classes were presented in this ontology. All these classes have been defined or extended in this work.

\begin{figure*}[h]
\centerline{
\includegraphics[width=1.1\textwidth]{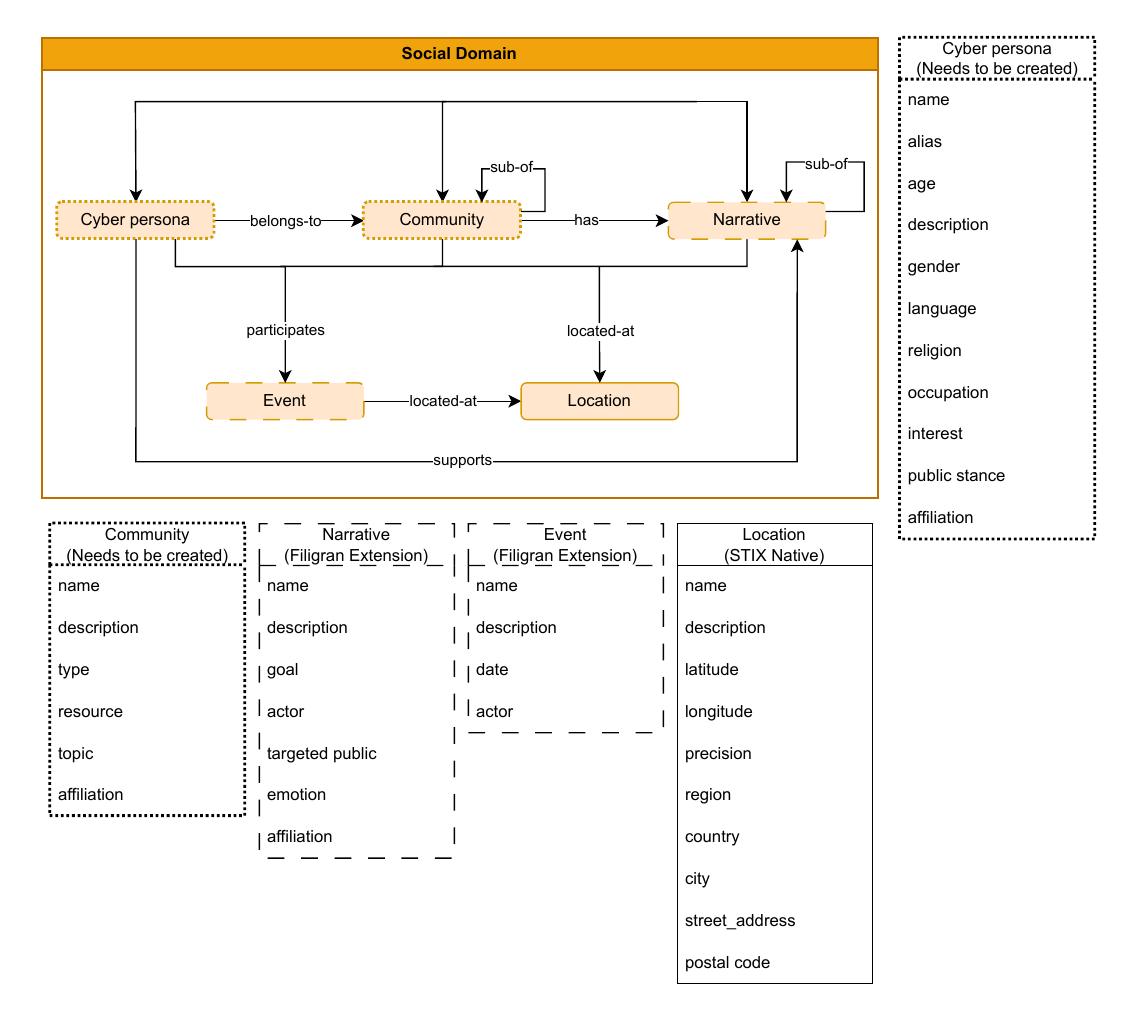}
}
\caption{Social domain visual representation}
\label{fig:social-domain}
\end{figure*}

\subsubsection{\texttt{Cyber Persona}}


A \texttt{Cyber Persona} is the virtual identity of the real (or assumed) person who manages one or more \texttt{User Accounts}. \texttt{Cyber Persona} is characterized by the name, description, occupation, public stance, and affiliation. The \texttt{Threat Actors} will try to shape, modify, or alter the perception, behaviors, or thoughts of the \texttt{Cyber Persona} using its human aspects like preferences, affiliations, or ideological tendencies. The attributes defined for \texttt{Cyber Persona} are described in \ref{appendix} (Table \ref{tab:cyber_persona}).

\subsubsection{\texttt{Community}}
 
\texttt{Communities} are groups of \texttt{Cyber Personas} who share values, interests or opinions. These communities influence perception, behavior, and decision-making, reinforcing worldviews and shaping social dynamics~\cite{LikeWar}. A \texttt{Community} comprises a name, description, type, and affiliation. Attacking, engaging, or modifying a community's behavior can be a key objective for any influencing operation because it affects all its \texttt{Cyber Personas} (mass impact). The attributes defined for \texttt{Community} are described in \ref{appendix} (Table \ref{tab:community}).


\subsubsection{\texttt{Narrative}}

A \texttt{Narrative} is a structured and coherent sequence of ideas, beliefs, or messages that influence how \texttt{Cyber Personas} or \texttt{Communities} perceive and interpret events, actions, or ideas, i.e., they provide context and meaning to what happens~\cite{LikeWar}. They reinforce views, ideologies, or programs that \texttt{Cyber Personas} \textit{support}, resulting in \texttt{Communities} \textit{having} certain \texttt{Narratives} that support their interests. In the context of IOs, \texttt{Threat actors} use the \texttt{Narratives} to encapsulate their actions and influence by producing different effects, such as guiding public opinion, legitimizing actions, or encouraging a community to act~\cite{MarshalGanz,LikeWar}. The attributes defined for \texttt{Narrative} are described in \ref{appendix} (Table \ref{tab:narrative}).

\subsubsection{\texttt{Event}}
An \texttt{Event} is real world occurrence such as elections, public shows, anniversaries, or other significant happenings that provide the broader context in which incidents may unfold or exert influence~\cite{filigran}. An \texttt{Event} is composed of a name, a description, and a date. These attributes are described in \ref{appendix} (Table \ref{tab:event}).

\subsubsection{\texttt{Location}}
The \texttt{Location} represents the geographic place where an \texttt{Event} or an \texttt{Entity} is \textit{located}. It consists of a name, a description, and a way of expressing a geographic point, such as coordinates or an address. The attributes defined for \texttt{Location} are described in \ref{appendix} (Table \ref{tab:location}).

\section{Conclusions and future work}






IOs are a complex problem that nations are trying to address. Disinformation, misinformation, and social polarization have become problems that affect democratic nations~\cite{Pamment2020}. It is necessary to have tools to identify, characterize, and adequately address these Influence Operations and their consequences~\cite{2eeas_report}. This paper presents an ontology for modeling IOs in the information environment. It allows a multidisciplinary approach to characterize the identified domains involved in IOs. It is worth mentioning that it maintains the main focus on CTI sharing as a basis for information sharing, providing a solution compatible with the most important CTI platforms. 

This work is currently under development with several lines of work underway:

\begin{itemize} 
\item The ontology has not yet been formally verified. It is still working to translate the actual description to the RDF formal language to validate correctly with tools like OOPS!~\cite{OOPS!}.

\item After the correct validation, a complete workflow will be developed to generate knowledge automatically. For this purpose, individuals will be automatically extracted from a dataset, processed with the ontology structure, and finally represented in a knowledge graph.

\item Similarly to other ontologies and CTI languages is necessary to develop a unified language for attributes that could be standardized as resource level in the Threat Actor class or Channel Type or Platform in the Channel class. Limiting the values for these attributes to a constrained list of options greatly facilitates the standardization and ease of sharing and understanding information unequivocally.

\end{itemize}

%% file: appendix.tex
\section{Tables of attributes of the Influence Operation Ontology classes}

\begin{table}[h]
    \centering
    \resizebox{\textwidth}{!}{%
    \begin{tabularx}{\textwidth}{|>{\raggedright\arraybackslash}p{2.5cm}|>{\raggedright\arraybackslash}p{1.2cm}|X|}
        \hline
        \multicolumn{3}{|c|}{\textbf{Incident Class}} \\ \hline
        \textbf{Attribute Name} & \textbf{Type} & \textbf{Description} \\ \hline
        Name (Mandatory) & String & A name to identify the incident \\ \hline
        Description & String & A description of the incident providing more context and details, usually including its purpose and what happened \\ \hline
        First Seen & Date & The first time that the incident was observed \\ \hline
        Last Seen & Date & The last time that the incident was observed \\ \hline
        Objective & String & The primary goal, objective, desired outcome, or intended effect of the incident \\ \hline
    \end{tabularx}
    }
    \caption{Incident Attributes. Source: \cite{stix,filigran}}
    \label{tab:incident}
\end{table}

\begin{table}[h]
    \centering
    \renewcommand{\arraystretch}{1.3}
    \resizebox{\textwidth}{!}{%
    \begin{tabularx}{\textwidth}{|>{\raggedright\arraybackslash}p{2.5cm}|>{\raggedright\arraybackslash}p{1.2cm}|X|}
        \hline
        \multicolumn{3}{|c|}{\textbf{Attack Pattern Class}} \\ \hline
        \textbf{Attribute Name} & \textbf{Type} & \textbf{Description} \\ \hline
        Name (Mandatory) & String & A name to identify the attack pattern \\ \hline
        Description & String & A description of the attack pattern providing more context and details, potentially including its purpose and key characteristics \\ \hline
        Alias & String (List) & Alternative names used to identify the attack pattern \\ \hline
        External Reference & String & The technique of the attack pattern from the DISARM Framework~\cite{disarm_framework} \\ \hline
        Kill Chain Phase & String & The tactic of the attack pattern from the DISARM Framework~\cite{disarm_framework} or the phase of Disarm Kill Chain~\cite{filigran} \\ \hline
    \end{tabularx}
    }
    \caption{Attack Pattern Attributes. Source: \cite{stix}}
    \label{tab:attack_pattern}
\end{table}
 
\begin{table}[h]
    \centering
    \renewcommand{\arraystretch}{1.2} 
    \resizebox{\textwidth}{!}{ 
        \begin{tabularx}{17cm}{|>{\raggedright\arraybackslash}p{2.5cm}|>{\raggedright\arraybackslash}p{2cm}|X|}
            \hline
            \multicolumn{3}{|c|}{\textbf{Campaign Class}} \\ \hline
            \textbf{Attribute Name} & \textbf{Type} & \textbf{Description} \\ \hline
            Name (Mandatory) & String & A name used to identify the campaign. \\ \hline
            Description & String & A description of the Campaign providing more context and details, usually including its purpose and its key characteristics. \\ \hline
            Aliases & String (List) & A list of other names that identify (or are believed to identify) this threat actor. \\ \hline
            First Seen & Date & The first time that the campaign was seen. \\ \hline
            Last Seen & Date & The last time that the campaign was seen. \\ \hline
            Objective & String & The primary goal, objective, desired outcome, or intended effect. That is, what the threat actor or intrusion set wants to achieve with this campaign. \\ \hline
        \end{tabularx}
    }
    \caption{Campaign Attributes. Source: \cite{stix}}
    \label{tab:campaign}
\end{table}

\begin{table}[h]
    \centering
    \renewcommand{\arraystretch}{1.2} 
    \resizebox{\textwidth}{!}{ 
        \begin{tabularx}{15.35cm}{|>{\raggedright\arraybackslash}p{2.5cm}|>{\raggedright\arraybackslash}p{2.5cm}|X|}
            \hline
            \multicolumn{3}{|c|}{\textbf{Threat Actor Class}} \\ \hline
            \textbf{Attribute Name} & \textbf{Type} & \textbf{Description} \\ \hline
            Name (Mandatory) & String & A name to identify the threat actor or threat actor group. \\ \hline
            Description & String & A detailed description of the threat actor, typically including its purpose and key characteristics. \\ \hline
            Threat Actor Type & Enumerated & The classification of this threat actor, such as cybercriminals, state-sponsored groups, hacktivists, or insiders. \\ \hline
            Aliases & String (List) & Alternative names used to identify this threat actor. \\ \hline
            First Seen & Date & The earliest known appearance or activity of the threat actor. \\ \hline
            Last Seen & Date & The most recent known activity of the threat actor. \\ \hline
            Roles & Enumerated & The different roles the threat actor may assume, such as activists, proxies, crime syndicates, or nation-states (not mutually exclusive). \\ \hline
            Goals & String & The primary objectives or intended outcomes of the threat actor’s activities. \\ \hline
            Sophistication & Enumerated & The skill level, expertise, and technical knowledge required to execute attacks. \\ \hline
            Resource Level & Enumerated & The level of organizational support and resources available to the threat actor. \\ \hline
            Primary Motivations & String & The main drivers behind the threat actor’s actions, defining their overall goals. \\ \hline
            Secondary Motivations & String & Additional factors influencing the threat actor, complementing but not replacing the primary motivation. \\ \hline
            Personal Motivations & String & Individual reasons driving a threat actor’s actions, which may align with or diverge from organizational objectives. \\ \hline
        \end{tabularx}
    }
    \caption{Threat Actor Attributes. Source: \cite{stix}}
    \label{tab:threat_actor}
\end{table}


\begin{table}[h]
    \centering
    \renewcommand{\arraystretch}{1.2}
    \resizebox{\textwidth}{!}{%
    \begin{tabularx}{\textwidth}{|>{\raggedright\arraybackslash}p{3.5cm}|>{\raggedright\arraybackslash}p{2.5cm}|X|}
        \hline
        \multicolumn{3}{|c|}{\textbf{Channel Class}} \\ \hline
        \textbf{Attribute Name} & \textbf{Type} & \textbf{Description} \\ \hline
        Name (Mandatory) & String & A name to identify the Channel. \\ \hline
        Description & String & A detailed explanation of the Channel, including its function, relevance, and role in message dissemination. \\ \hline
        Platform & Enumerated & The kind of platform or medium used. \\ \hline
        Affiliation & String & Any known connection to organizations, networks, or influence operations. \\ \hline
        Reach & String & The estimated audience size and engagement level of the Channel. \\ \hline
        Purpose & String & The primary goal of the Channel. \\ \hline
        Sponsored & Boolean & Whether the Channel is financially supported or promoted by an external entity, such as a government or private organization. \\ \hline
        Channel Type & Enumerated & The classification of the channel based on its affiliation and control structure (e.g official Communication Channel, State-linked channels, State-controlled Channels). \\ \hline
    \end{tabularx}
    }
    \caption{Channel Attributes}
    \label{tab:channel}
\end{table}

\begin{table}[h]
    \centering
    \renewcommand{\arraystretch}{1.2}
    \resizebox{\textwidth}{!}{%
    \begin{tabularx}{\textwidth}{|>{\raggedright\arraybackslash}p{2.5cm}|>{\raggedright\arraybackslash}p{2.5cm}|X|}
        \hline
        \multicolumn{3}{|c|}{\textbf{Message Class}} \\ \hline
        \textbf{Attribute Name} & \textbf{Type} & \textbf{Description} \\ \hline
        Name (Mandatory) & String & A name to identify the Message. \\ \hline
        Description & String & A detailed message explanation, including its purpose, main topic, and relevant contextual information. \\ \hline
        Media Content & MediaObject (schema.org) & The multimedia elements associated with the Message, such as text, images, videos, or audio. \\ \hline
        URL & String & The web address where the Message is hosted or published. \\ \hline
        Format & String & The type of Message, e.g.,  article, video, news report, or social media post. \\ \hline
    \end{tabularx}
    }
    \caption{Message attributes}
    \label{tab:message}
\end{table}

\begin{table}[h]
    \centering
    \renewcommand{\arraystretch}{1.2} 
    \resizebox{\textwidth}{!}{ 
        \begin{tabularx}{15cm}{|>{\raggedright\arraybackslash}p{2.5cm}|>{\raggedright\arraybackslash}p{2.5cm}|X|}
        \hline
        \multicolumn{3}{|c|}{\textbf{User Account Class}} \\ \hline
        \textbf{Attribute Name} & \textbf{Type} & \textbf{Description} \\ \hline
        Display Name (Mandatory) & String & The nickname of the user account, which may differ from the actual name. \\ \hline
        Name  & String & The real or chosen name associated with the account. \\ \hline
        Age & Integer & The age of the account owner. \\ \hline
        Icon & Image (schema.org) & The profile image or avatar representing the account. \\ \hline
        Description & String & A brief bio or tagline summarizing the user’s purpose, interests, or affiliations. \\ \hline
        External Links & String (List) & URLs linking to other profiles, websites, or resources related to the account. \\ \hline
        Region & String & The geographical location associated with the account. \\ \hline
        Account Created & Date & The date when the account was registered on the platform. \\ \hline
        Platform & String & The specific internet service or social media network where the account operates. \\ \hline
        Privacy Settings & String & The level of visibility (public, private, restricted) set by the user. \\ \hline
        Followers & Integer & The number of users who subscribe to or follow the account’s updates. \\ \hline
        Following & Integer & The number of other accounts this user follows. \\ \hline
        Rating & Integer & A metric that reflects the account's reputation, trust, or engagement level. \\ \hline
        Privileged & Boolean & Indicates whether the account has special privileges (e.g., verified status, admin rights). \\ \hline
        Disabled & Boolean & Specifies whether the account is active, suspended, or permanently banned. \\ \hline
        Automation & Integer & A number that indicates the automatic actions' level. \\ \hline
    \end{tabularx}
    }
    \caption{User Account attributes \cite{stix}}
    \label{tab:user_account}
\end{table}

 

\begin{table}[h]
    \centering
    \renewcommand{\arraystretch}{1.2} 
    \resizebox{\textwidth}{!}{ 
        \begin{tabularx}{14cm}{|>{\raggedright\arraybackslash}p{2.5cm}|>{\raggedright\arraybackslash}p{1.2cm}|X|}
        \hline
        \multicolumn{3}{|c|}{\textbf{Cyber Persona Class}} \\ \hline
        \textbf{Attribute Name} & \textbf{Type} & \textbf{Description} \\ \hline
        Name (Mandatory) & String & The real name of the cyber persona \\ \hline
        Alias & String & Alternative names, usernames, or pseudonyms the cyber persona uses in digital environments \\ \hline
        Age & Integer & The age of the cyber persona, whether real or self-reported \\ \hline
        Description & String & A brief bio summarizing the cyber persona’s physical characteristics, interests, affiliations, or online presence \\ \hline
        Gender & String & Gender with which the cyber persona identifies itself \\ \hline
        Language & String & The primary language(s) used by the cyber persona \\ \hline
        Religion & String & The religious beliefs or ideologies associated with the cyber persona \\ \hline
        Occupation & String & The profession, role, or function of the cyber persona \\ \hline
        Interest & String & Topics, activities, or fields of engagement of the cyber persona \\ \hline
        Public Opinion & String & Expressed viewpoints, perspectives, or ideological positions that the cyber persona actively shares or supports online \\ \hline
        Affiliation & String & Any known connection to organizations, networks, or influence operations \\ \hline
    \end{tabularx}}
    \caption{Cyber Persona attributes}
    \label{tab:cyber_persona}
\end{table}

\begin{table}[h]
    \centering
    \renewcommand{\arraystretch}{1.2} 
    \begin{tabularx}{\textwidth}{|>{\raggedright\arraybackslash}p{2.5cm}|>{\raggedright\arraybackslash}p{2.2cm}|X|}
        \hline
        \multicolumn{3}{|c|}{\textbf{Community Class}} \\ \hline
        \textbf{Attribute Name} & \textbf{Type} & \textbf{Description} \\ \hline
        Name (Mandatory) & String & A name to identify the Community \\ \hline
        Description & String & A brief overview of the community, its purpose, and its main characteristics. \\ \hline
        Type & String & The nature of the group, e.g., organization, corporate, social, ideological \\ \hline
        Resources & Enumerated & The assets the community relies on, such as members, funding, information, or technology \\ \hline
        Topic & String & The central themes or issues that unite the community \\ \hline
        Affiliation & String & Any known connection to organizations, networks, or influence operations \\ \hline
    \end{tabularx}
    \caption{Community attributes}
    \label{tab:community}
\end{table}

\begin{table}[h]
    \centering
    \renewcommand{\arraystretch}{1.2} 
    \begin{tabularx}{\textwidth}{|>{\raggedright\arraybackslash}p{2.5cm}|>{\raggedright\arraybackslash}p{1.2cm}|X|}
        \hline
        \multicolumn{3}{|c|}{\textbf{Narrative Class}} \\ \hline
        \textbf{Attribute Name} & \textbf{Type} & \textbf{Description} \\ \hline
        Name (Mandatory) & String & A name to identify the Narrative \\ \hline
        Description & String & A summary of the narrative's core content, context, and variations \\ \hline
        Goal & String & The main purpose of the narrative, what this narrative is trying to produce or obtain \\ \hline
        Topic & String & The central theme (politics, economy, health, security, etc.) \\ \hline
        Targeted Public & String & The intended audience based on demographics, psychographics, and online communities \\ \hline
        Emotion & String & The main emotions triggered by the narrative \\ \hline
        Affiliation & String & Any known connection to organizations, networks, or influence operations \\ \hline
    \end{tabularx}
    \caption{ Narrative attributes. Source \cite{filigran}.}
    \label{tab:narrative}
\end{table}

\begin{table}[h]
    \centering
    \renewcommand{\arraystretch}{1.2} 
    \begin{tabularx}{\textwidth}{|>{\raggedright\arraybackslash}p{2.5cm}|>{\raggedright\arraybackslash}p{1.2cm}|X|}
        \hline
        \multicolumn{3}{|c|}{\textbf{Location Class}} \\ \hline
        \textbf{Attribute Name} & \textbf{Type} & \textbf{Description} \\ \hline
        Name (Mandatory) & String & A name used to identify the location \\ \hline
        Description & String & A textual description of the location \\ \hline
        Latitude & Float & The location's latitude in decimal degrees, where positive values indicate positions north of the equator and negative values represent positions south of the equator \\ \hline
        Longitude & Float & The location's longitude in decimal degrees, where positive values indicate longitudes east of the prime meridian and negative values indicate longitudes west of the prime meridian \\ \hline
        Precision & String & Defines the precision of the coordinates specified by the latitude and longitude properties \\ \hline
        Region & String & The region that this location describes \\ \hline
        Country & String & The country that this location describes \\ \hline
        City & String & The city that this location describes \\ \hline
        Street Address & String & The street address that this location describes \\ \hline
        Postal Code & String & The postal code for this location \\ \hline
    \end{tabularx}
    \caption{Location Attributes. Source: \cite{stix}}
    \label{tab:location}
\end{table}

\begin{table}[h]
    \centering
    \renewcommand{\arraystretch}{1.2}
    \resizebox{\textwidth}{!}{%
    \begin{tabularx}{\textwidth}{|>{\raggedright\arraybackslash}p{2.5cm}|>{\raggedright\arraybackslash}p{1.2cm}|X|}
        \hline
        \multicolumn{3}{|c|}{\textbf{Event Class}} \\ \hline
        \textbf{Attribute Name} & \textbf{Type} & \textbf{Description} \\ \hline
        Name (Mandatory) & String & A name to identify the event. \\ \hline
        Description & String & A description of the event providing more context and details, potentially including its purpose and key characteristics like the topic and actors. \\ \hline
        Date & Date & When the event was or will take place. \\ \hline
    \end{tabularx}
    }
    \caption{Event Attributes. Source: \cite{filigran,stix}}
    \label{tab:event}
\end{table}